\documentclass[aps,prl,twocolumn,showpacs]{revtex4}
\usepackage{graphicx}

\begin{document}
\title{The quantized Hall effect in the presence of resistance 
fluctuations
}
\author{E. Peled}
\author{D. Shahar}
\affiliation{
Department of Condensed Matter Physics, Weizmann Institute, Rehovot 
76100, Israel
}
\author{Y. Chen}
\affiliation{
Department of Electrical Engineering, Princeton University, 
Princeton, New Jersey 08544, USA
}
\author{D. L. Sivco}
\author{A. Y. Cho}
\affiliation{
Bell Laboratories, Lucent Technologies, 600 Mountain Avenue, Murray 
Hill, 
New Jersey 07974, USA
}
\date{\today}

\begin{abstract}
We present an experimental study of 
mesoscopic, two-dimensional electronic systems at high magnetic fields.
Our samples, prepared from a low-mobility
InGaAs/InAlAs wafer, exhibit reproducible, sample specific, 
resistance fluctuations. 
Focusing on the lowest Landau level we find that, 
while the diagonal resistivity
displays strong fluctuations, the Hall resistivity 
is free of fluctuations and remains quantized at its $\nu=1$ value, 
$h/e^{2}$. This is true also in the insulating phase that 
terminates the quantum Hall series. These results extend the validity 
of the semicircle law of conductivity 
in the quantum Hall effect to the mesoscopic regime.
\end{abstract}

\pacs{73.43.-f, 71.30.+h, 72.80.Sk, 73.23.-b}

\maketitle


For small, mesoscopic samples, the conductivity in the Quantum Hall 
(QH)
regime exhibits reproducible, sample specific, fluctuations
\cite{Chang1988SSC67,Timp1987PRL59,Hohls2002PRB66,
Cobden1999PRL82,Machida2001PRB63,
Main1994PRB50,Simmons1989PRL63,Goldman1995SC267,
Cobden1996PRB54,Cho1997PRB55,Ando1994PRB49,Wang1996PRL77}.  
Although quite interesting by themselves, the greatest benefit in 
studying these fluctuations is that they can provide clues to the 
nature of the 
mechanism responsible for conduction in the QH regime. 

There are several known
mechanisms that can lead to such fluctuations.  
Fluctuations as a result of the modification of the 
interference between many possible electron paths across the sample 
are at
the heart of the theory of Universal Conductance Fluctuations (UCF)
\cite{Lee1987PRB35}.  Although this theory is not expected to be 
valid at
high magnetic fields ($B$), modified UCF theories have been suggested
\cite{Xiong1992PRL68,Maslov1993PRL71} to take into account the 
influence of
$B$ on the electron trajectories, and several experiments were
interpreted in terms of UCF at high $B$ 
\cite{Timp1987PRL59,Hohls2002PRB66}. 

A different point of view has been suggested by two recent 
experimental studies: Cobden {\it et al.} \cite{Cobden1999PRL82}, and 
Machida {\it et al.}
\cite{Machida2001PRB63} have found that in the QH regime fluctuation 
patterns follow straight lines in the magnetic field--carrier-density plane. 
These lines appear to be parallel to integer 
Landau level (LL) filling-factor lines.  These results were 
attributed to charging of electron
puddles in the sample \cite{Cobden1999PRL82}, or to changes in a
compressible-strip network configuration \cite{Machida2001PRB63}.  
 
Another possible source of fluctuations, expected to dominate when 
$B$ is high enough such that only a small number of  LL are occupied, 
is
resonant tunneling \cite{Jain1988PRL60}.  In this process fluctuations
arise when an electron scatters from one edge of the sample to the 
other
through a bulk impurity.  This model was found to be consistent with
observed high-$B$ fluctuations \cite{Main1994PRB50}, and has been 
used in measurements of fractional 
charge \cite{Simmons1989PRL63,Goldman1995SC267}. 
Applying this model to the case when only 
the lowest LL is occupied, Jain and Kivelson 
\cite{Jain1988PRL60} argued that the fluctuations will be limited to 
the diagonal resistivity, $\rho_{xx}$, leaving the Hall resistivity, 
$\rho_{xy}$, quantized. This conjecture was contended by 
B\"{u}tikker \cite{Buttiker1989PRL62}.  
Experiments have so far shown that fluctuations in $\rho_{xx}$ are 
accompanied by fluctuations in $\rho_{xy}$ 
\cite{Chang1988SSC67,Timp1987PRL59,Main1994PRB50}, 
although these experiments were limited to relatively lower $B$'s, 
where 
more than one  LL participates in the conduction ($\nu \geq 2$). 

In this letter we report on an experimental study of the resistivity 
of
mesoscopic samples in the QH regime, focusing mainly on the lowest LL.
We find that, while $\rho_{xx}$ may exhibit large fluctuations,
$\rho_{xy}$ remains nearly fluctuation-free and quantized at its 
$\nu=1$ QH
value, $h/e^{2}$.  This is true even beyond the transition to the
insulating phase that terminates the QH series. 

Our data were obtained from two samples, T2Ga and T2C, wet-etched 
from the
same InGaAs/InAlAs wafer that contained, after illumination with an 
LED, a two-dimensional electronic system
in a 200 \AA\ quantum well.  We defined a Hall-bar geometry with
lithographic widths of $W= 10\ \mu$m for sample T2Ga and $W=2\ \mu$m 
for
sample T2C, and voltage-probe separation of $L=2\times W$, 
maintaining an identical aspect ratio.  
The samples were cooled in a dilution refrigerator with
a base $T$ of 12 mK.  The mobility and density of sample
T2Ga were 
$\mu = 22,000$ cm$^{2}$/Vsec and $n_{s}=1.5 \cdot 10^{11}$ cm$^{-2}$.
Sample T2C was cooled down twice, with
$\mu= 23,850$ cm$^{2}$/Vsec, $n_{s}=1.83 \cdot 10^{11}$ cm$^{-2}$ and
$\mu= 13,700$ cm$^{2}$/Vsec, $n_{s}=1 \cdot 10^{11}$ cm$^{-2}$ at the 
first (T2Cc) and second (T2Ci) cool-downs, respectively.
Four-probe measurements were done using standard AC
lock-in techniques with excitation currents of 0.1 -- 1 nA and 
frequencies
of 1.7 -- 11 Hz.  Temperatures below 30 mK are nominal, since the
resistivity does not change in the $T$ range 12 -- 30 mK.

\begin{figure}[ht]
\includegraphics[scale=0.55]{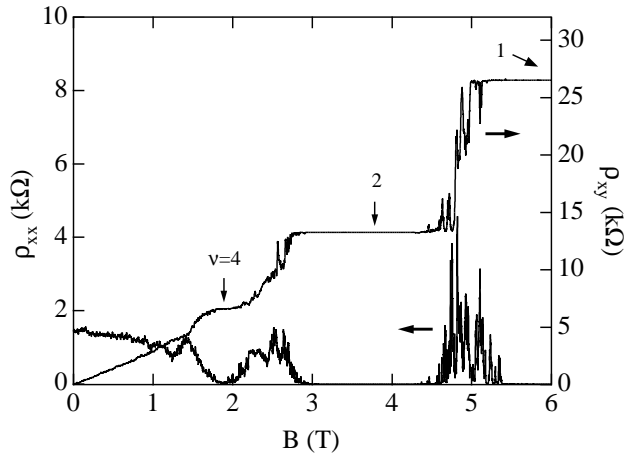} 
\caption{
$\rho_{xx}$ and $\rho_{xy}$ vs. $B$ for sample T2Cc, $T=$ 12 mK. 
QH plateaus are designated by their filling factor.
}
\label{fig:Fig1_T2Cc2_QHE}
\end{figure}

In Fig.\ \ref{fig:Fig1_T2Cc2_QHE} we plot $\rho_{xx}$ and 
$\rho_{xy}$ for sample T2Cc, taken over a broad range of $B$. 
Although the sample is relatively small ($W=2\ \mu$m), integer QH 
plateaus in 
$\rho_{xy}$ (designated by their filling factor) and the associated minima 
in $\rho_{xx}$ are clearly observed. Due to the low mobility of our 
samples, fractional QH features are not seen down to the lowest $T$.
In addition to the QH features, $\rho_{xx}$ and $\rho_{xy}$ exhibit
reproducible fluctuations, which maintain their pattern as long as 
the sample is kept cold, attesting to the mesoscopic nature of our samples.
The fluctuations decrease in magnitude when $T$ is increased and have 
a similar pattern when switching to voltage probes that are on the opposite 
side of the Hall-bar (for $\rho_{xx}$ measurements)
or to a neighbouring pair of voltage probes (for $\rho_{xy}$ 
measurements).
At the low-$B$ range, we associate the fluctuations with the theory 
of 
UCF. According to this theory,  the coherence length ($L_{\phi}$) can 
be deduced from the amplitude and correlation-$B$ of the 
fluctuations. $L_{\phi}$ in our samples is found to be between 1.2 to 3 
$\mu$m at base $T$, varying from sample to sample and 
between cooldowns.
The larger sample, T2Ga, also exhibits reproducible fluctuations, but 
of reduced amplitude, due to averaging of subunits of size $L_{\phi}$ 
over the sample area.
At $T=$ 12 mK and near $B=0$ the rms of the fluctuations in samples T2Cc 
and T2Ga is 35 and 9.6 
$\Omega$, respectively, corresponding to conductivity fluctuations 
$\delta\sigma=0.54$ and 0.067 $e^2/h$. As expected, the reproducible 
fluctuations are not limited to 
the low-$B$ range of Fig.\ \ref{fig:Fig1_T2Cc2_QHE}, and are present 
over 
the entire $B$ range of our measurements.  
At the high-$B$ range the fluctuations appear on top of the
usual QH features, both in $\rho_{xx}$ and in $\rho_{xy}$, as seen 
in 
previous studies \cite{Chang1988SSC67,Timp1987PRL59, Hohls2002PRB66,
Cobden1999PRL82, Machida2001PRB63,
Main1994PRB50,Simmons1989PRL63, Goldman1995SC267,
Cobden1996PRB54}.

\begin{figure}[ht]
\includegraphics[scale=0.55]{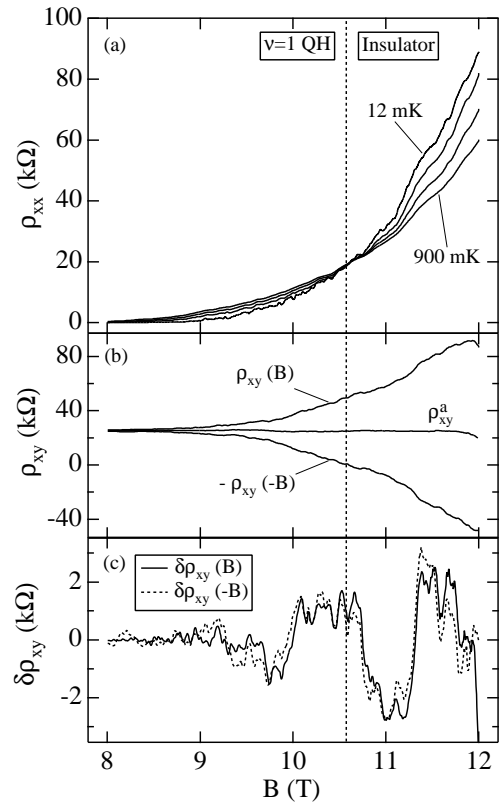} 
\caption{The resistivity of sample T2Ga at 
$\nu<1$. 
(a) $\rho_{xx}$ traces. Temperatures are 
12, 500, 700, and 900 mK. The crossing point of 
the traces is at $B_{c}=$ 10.62 T ($\nu_{c}=$ 0.58), indicated by the 
dotted vertical line. 
(b) $\rho_{xy}(B)$, $-\rho_{xy}(-B)$, and the antisymmetric part, 
$\rho_{xy}^{a}$, at $T=$ 16 mK. 
(c) Comparison of the $\pm B$ fluctuations of $\rho_{xy}$. 
The fluctuation patterns 
are very similar, and overlap to within 92\%.
This symmetry 
leads to much smaller fluctuations in $\rho_{xy}^{a}$.
}
\label{fig:Fig2_T2Ga_RxxRxy}
\end{figure}

In the remainder of this letter we will focus on data taken at 
$\nu<1$, where only the lowest LL contributes to the transport.
In Fig.\ \ref{fig:Fig2_T2Ga_RxxRxy} we examine the resistivity 
obtained from 
sample T2Ga, in the $B$ range of 8--12 T, corresponding to $\nu=$ 
0.775--0.517. 
We begin by focusing on $\rho_{xx}$ traces taken at $T=$12,
500, 700, and 900 mK in Fig.\ \ref{fig:Fig2_T2Ga_RxxRxy}(a).
Two different transport regimes can be distinguished according
to the $T$-dependence of $\rho_{xx}$: the QH liquid regime (low-$B$
side), where $\rho_{xx}$ increases with increasing $T$, and the 
insulating regime,
(high-$B$ side), where $\rho_{xx}$ decreases with $T$.  The 
transition $B$, $B_{c}=$ 10.62 T ($\nu_{c}=0.58$), where  $\rho_{xx}$ 
is $T$-independent, is indicated by the dotted vertical line.  
The value of $\rho_{xx}$ at the transition, $\rho_{xxc}=0.77\ 
h/e^{2}$, 
is in agreement with previous
experimental results where $\rho_{xxc}\sim h/e^{2}$ 
\cite{Hughes1994JPCM6,
Shahar1995PRB52, Shahar1995PRL74, Dunford2000PE6}.
The $\rho_{xx}$ fluctuations here are much larger 
than at low-$B$, and have an amplitude as high as 2,840 $\Omega$ 
on the insulating side.

Before we proceed with the main result of our work, which stems from 
the 
analysis of the $\rho_{xy}$ data, a word of caution is in order. An 
experimental 
determination of $\rho_{xy}$ in the lowest LL
is difficult, because an admixture of $\rho_{xx}$ into the data is
unavoidable.  This is especially important in the insulating regime 
where $\rho_{xx}$ is large, 
and even a small fraction of $\rho_{xx}$ admixture can result in a 
significant change in the measured $\rho_{xy}$.  The 
traditional way to overcome this problem is by using the $B$-symmetry
of the resistivity components: $\rho_{xx}$ is expected to be symmetric
in $B$, while $\rho_{xy}$ is antisymmetric.  The
true $\rho_{xy}$ is obtained, in principle, by an antisymmetrization 
of 
the measured  $\rho_{xy}(B)$ and $\rho_{xy}(-B)$:
$\rho_{xy}^{a} =\frac{1}{2}[\rho_{xy}(B)-\rho_{xy}(-B)]$.  In 
practice, 
because the admixture is a result of both contact
misalignment and current nonuniformities in the sample, $\rho_{xx}$ 
itself is not entirely 
$B$-symmetric, limiting the accuracy with which $\rho_{xy}$ can be 
determined. 

We now turn to the analysis of the $\rho_{xy}$ data at $\nu<1$. In 
Fig.\ \ref{fig:Fig2_T2Ga_RxxRxy}(b) we plot $\rho_{xy}(B)$, 
$-\rho_{xy}(-B)$ and the antisymmetrized $\rho_{xy}^{a} $ taken at 
$T=$ 16 mK. As in previous works 
\cite{Hilke1999EL46,Hilke1998Nature395,
Dunford2000PE6,Lang2001CM0106375}, we find that $\rho_{xy}(B)$ and
$\rho_{xy}(-B)$ show a strong, symmetric, $B$-dependence that is 
canceled out of $\rho_{xy}^{a}$, resulting in a Hall coefficient that 
remains nearly constant 
and quantized at its $\nu=1$ value, $h/e^{2}$. This holds into the 
insulating phase, 
which has been termed the quantized Hall insulator (QHI) 
\cite{Hilke1998Nature395}.
The novelty in our data is that, while $\rho_{xy}(\pm B)$ include 
fluctuations, they are symmetric in $B$ and therefore do not appear in 
$\rho_{xy}^{a}$.
To demonstrate this symmetry more clearly, we compare in
Fig.\ \ref{fig:Fig2_T2Ga_RxxRxy}(c) the positive and negative $B$ 
fluctuation patterns, $\delta \rho_{xy}(\pm B)$, after subtracting a 
smooth 
background from the original traces. $\delta \rho_{xy}(\pm B)$ are 
very
similar in shape and magnitude, and overlap to within 92\%
\cite{OverlapNote}. The maximum amplitude of the 
$\rho_{xy}(\pm B)$ fluctuations is 2,825 $\Omega$, 
while that of $\rho_{xy}^{a}$ is 735 $\Omega$.
We attribute these remaining $\rho_{xy}$ fluctuations to an 
asymmetric component of $\rho_{xx}$ in sample T2Ga. Measurements of 
$\rho_{xx}(\pm B)$ reveal an asymmetric component whose magnitude is 
consistent with the fluctuations of $\rho_{xy}^{a}$. We stress that 
even if sample inhomogeneities were not present, obtaining perfectly 
reproducible resistivity measurements at opposite $B$ polarities is 
not practical due to the stringent requirements this will place on 
$T$ stability when the $B$ field is swept over nearly 20 T. This 
problem becomes even more severe in light of the strong 
$T$-dependence of $\rho_{xx}$ in the insulating phase.

\begin{figure}[ht]
\includegraphics[scale=0.55]{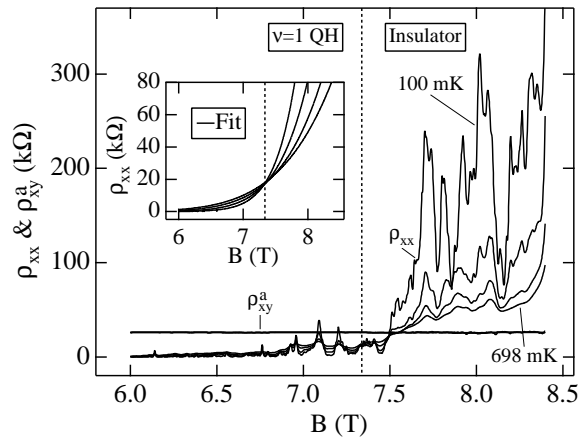} 
\caption{$\rho_{xx}$ and $\rho_{xy}^{a}$ of sample T2Ci at high $B$. 
$T=$ 100, 
295, 500, and 698 mK. 
The large $\rho_{xx}$ fluctuations make it 
difficult to identify the transition point. Inset: we use the 
crossing point of the smooth $\rho_{xx}$-fit curves \cite{FitNote}
to identify $B_{c}$: $B_{c}=$ 
7.34 T ($\nu_{c}=$ 0.572), $\rho_{xxc}=0.67\ h/e^{2}$.
}
\label{fig:Fig3_T2Ci_RxxRxy}
\end{figure}

In order to enhance the fluctuating part of the data, we repeated
our $\nu<1$ measurements with the smaller ($W=2\ \mu$m) sample, T2Ci, 
which was fabricated with utmost care to reduce contact misalignments 
to the minimum. 
The results are shown in Fig.\ \ref{fig:Fig3_T2Ci_RxxRxy}, where we 
plot $\rho_{xx}$ and $\rho_{xy}^{a}$ vs. $B$ taken at $T=$ 100, 295, 
500, and 698 
mK. We start by examining the $\rho_{xx}$ data, which now appear to 
be dominated by fluctuations that are as large as 188,000 $\Omega$ on 
the
insulating side.  Nevertheless, 
by fitting a smooth curve through the data, the average $\rho_{xx}$ 
can be determined and the QH and insulating regimes properly 
identified (see inset). The fitted curves cross 
at $B_{c}=$ 7.34 T ($\nu_{c}=0.572$) and $\rho_{xxc}=0.67\ h/e^{2}$. 

In stark contrast with the $\rho_{xx}$ data, $\rho_{xy}^{a}$ 
appears to be free of fluctuations. An upper-bound estimate of the 
magnitude of the $\rho_{xy}^{a}$ fluctuations is 160--220 
times less than 
the amplitude of $\delta\rho_{xx}$ at similar background values. 
Comparing these data to the data obtained from the 10 $\mu$m sample, 
we see that reducing the sample size by a factor of 5 resulted in an 
increase of $\delta\rho_{xx}$ by nearly 70 while 
$\delta\rho_{xy}$ increased by only a factor of 2. 

We now turn to a discussion of the consequences of our results.
So far we have shown that for $\nu<1$ the two resistivity components 
of our samples show
dissimilar behaviour: $\rho_{xx}$ displays large reproducible 
fluctuations
while $\rho_{xy}$ remains close to its quantized value and shows 
only insignificant fluctuations. 

Of the models mentioned above 
pertaining 
to fluctuations in the QH 
regime only the resonant-tunneling model 
specifically deals with the behaviour of $\rho_{xy}$.
Our measurements of a fluctuation-free, quantized, $\rho_{xy}$ are 
consistent with this 
model. 

\begin{figure}[ht]
\includegraphics[scale=0.5]{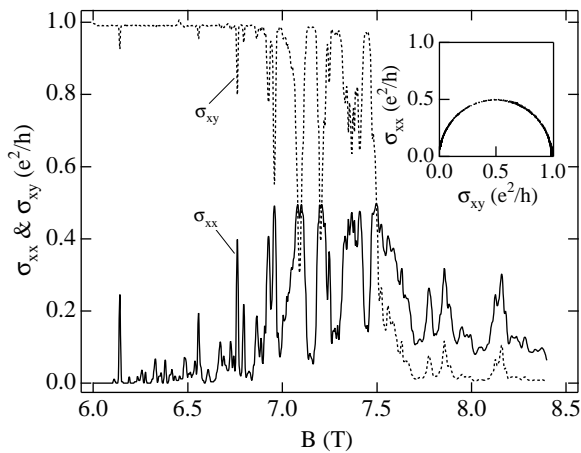} 
\caption{$\sigma_{xx}$ and $\sigma_{xy}$ of sample T2Ci near the 
QH-insulator 
transition, calculated by inverting the resistivity tensor. $T=$ 100 
mK. Inset: The semicircle relation, $\sigma_{xx}$ vs. $\sigma_{xy}$.
}
\label{fig:Fig4_Semicircle}
\end{figure} 

There are two additional models for conduction in the QH regime that 
predict a quantized $\rho_{xy}$, although they do not directly address the properties of the
mesoscopic fluctuations of the resistivity.
In the first model, Ruzin and his collaborators 
\cite{Dykhne1994PRB50,Ruzin1995PRL74} treat the electron system in 
the QH transition region as a random mixture of two phases, and
find a semicircle relation for the conductivity components, 
$\sigma_{xx}$ and $\sigma_{xy}$.
For the QH-insulator transition, this relation is:
\begin{displaymath}
     \sigma_{xx}^{2} + \left( \sigma_{xy} - 
     \frac{e^{2}}{2h}\right)^{2} = \left(\frac{e^{2}}{2h}\right)^{2}, 
 \end{displaymath}
which is mathematically equivalent to the quantization of 
$\rho_{xy}$  \cite{Hilke1998Nature395}. 
Since we are unable to measure the conductivity components directly, we obtain them  by inverting the resistivity tensor, $\rho_{xx}$ and $\rho_{xy}^{a}$.
In Fig.\ \ref{fig:Fig4_Semicircle} we plot the resulting $\sigma_{xx}$ and $\sigma_{xy}$
of sample T2Ci at the QH-insulator transition. In contrast with
 the resistivity tensor, both $\sigma_{xx}$ and $\sigma_{xy}$ display
 fluctuations, stemming from their mutual dependence on $\rho_{xx}$ as 
well as on $\rho_{xy}^{a}$. Nevertheless, $\sigma_{xx}$ and 
$\sigma_{xy}$ 
obey the semicircle relation, as shown in the inset of Fig.\ 
\ref{fig:Fig4_Semicircle}, indicating the special correlation that 
exists between the fluctuations of the conductivity-tensor components.
Our work extends the validity of the semicircle relation to the 
mesoscopic regime of transport.

Another model pertaining to the quantization of $\rho_{xy}$ near the 
transition, and in the insulating regime, has been proposed by 
Shimshoni and Auerbach \cite{Shimshoni1997PRB55}. They
showed that  
transport in a random Chalker-Coddington-based network of puddles produces a 
quantized $\rho_{xy}$ when $L_{\phi}$ is smaller than the puddle size. 
However, $\rho_{xy}$ is expected to diverge when 
$L_{\phi}$ is larger than the puddle size 
\cite{Pryadko1999PRL82, Zulicke2001PRB63}.
It would be interesting to see how this model can be extended to 
accommodate samples that exhibit large resistance fluctuations.

Finally, we wish to emphasize that the absence of fluctuations in 
$\rho_{xy}$ near the QH-insulator transition is not inconsistent with 
the results of previous experiments (and the present work, see Fig.\ 
\ref{fig:Fig1_T2Cc2_QHE}), which show large $\rho_{xy}$ fluctuations 
that survive the averaging of opposite $B$ directions in the vicinity 
of the transitions in higher LL's. We recall the 
mapping \cite{Kivelson1992PRB46} that exists between transitions at 
higher LL's and the `basic', QH-insulator, transition at the lowest 
LL: in this mapping one regards the higher transitions as a 
QH-insulator transition occurring in the presence of a number of full 
and inert LL's. A simple calculation shows that the $\rho_{xy}$ of 
higher LL transitions will include components proportional to 
$\rho_{xx}$ of a QH-insulator transition \cite{Shahar1997PRL79}.
We leave for future work a more direct 
verification of this mapping to the fluctuating part of the 
resistivity.
 
To conclude, we have shown that the quantization of the Hall effect 
in two-dimensional electron systems can be maintained in the 
mesoscopic regime, even when the diagonal resistivity, $\rho_{xx}$, 
is nonzero and exhibits large fluctuations. These results are in 
agreement with the predictions of Jain and 
Kivelson \cite{Jain1988PRL60} and with 
the semicircle relation for the conductivity components 
\cite{Dykhne1994PRB50,Ruzin1995PRL74}.
 
We Wish to thank A. Auerbach, Y. Oreg, E. Shimshoni, A. Stern, and
D. C. Tsui for useful discussions.
This work is supported by the BSF and by the Koshland Fund. Y. C. is supported by the (US) NSF.


\begin{thebibliography}{10}
\bibitem{OverlapNote}
We define the overlap as: \newline \( overlap = 100 \times
  cor(\delta \rho_{xy}(B),\delta \rho_{xy}(-B))/ \) \newline 
  \(\sqrt{cor(\delta \rho_{xy}(B),\delta \rho_{xy}(B))}/
  \sqrt{cor(\delta \rho_{xy}(-B),\delta
  \rho_{xy}(-B))}\), where $cor$ stands for the value of the
  correlation function at $\Delta B=0$.
\end{thebibliography}

\end{document}